\documentclass[12pt,a4paper]{article}
\usepackage[utf8]{inputenc}
\usepackage[T1]{fontenc}
\usepackage{amsmath}
\usepackage{amssymb}
\usepackage{graphicx}
\usepackage[numbers]{natbib}
\usepackage{geometry}
\title{Progress: A Post-AI Manifesto}
\author{Christoforus Yoga Haryanto}

\begin{document}

\maketitle

\begin{abstract}
This manifesto outlines key principles for progress in the post-AI era, emphasizing non-linear yet cumulative advancement, deep understanding of purpose and context, multi-stakeholder collaboration, and system-level experimentation. It redefines progress as substantial, durable, and replicable advancement, highlighting the importance of balancing technological innovation with human-centric values. It acknowledges AI's potential to accelerate progress across industries while recognizing its limitations, such as creating illusions of understanding and potentially narrowing problem-solving approaches. It concludes that true progress in the AI age requires a symbiosis of artificial intelligence capabilities and human ingenuity, calling for a holistic, interdisciplinary approach to shape a future that serves all of humanity.
\end{abstract}

\section{Key Stances}
To guide us in making progress in today's Post-AI society we have to understand that:
\begin{enumerate}
    \item Progress is non-linear but cumulative, with paradigm shifts and iterative processes.
    \item A deep understanding of purpose, mechanisms, and historical context is essential.
    \item Multi-stakeholder collaboration at disciplinary intersections is necessary for progress.
    \item Methodological system-level experimentation and prototyping drive innovation.
    \item Conflicting progress should be embraced as a tool for refinement and validation.
    \item The post-AI era requires a holistic, interdisciplinary approach to progress.
    \item AI can accelerate advances in multiple industries but has fundamental limitations.
    \item Human creativity, consciousness, and abductive reasoning are necessary in the AI era.
    \item Progress must balance technological advancement with human-centric values.
    \item True progress emerges from a symbiosis of AI capabilities and human ingenuity.
\end{enumerate}

We, the architects of tomorrow, declare these principles as the foundation of our shared vision. We reject the status quo that stifles innovation and perpetuates inequality. Instead, we embrace a future where progress serves humanity, not just the privileged.

\section{What is Progress?}
Progress: A substantial and durable advancement that builds upon previous achievements, characterized by its repeatability and potential to serve as a foundation for future developments. While progress may eventually be superseded or rendered obsolete by newer advancements, it often remains an integral part of the evolutionary chain of improvements. Progress is not necessarily labour-intensive and can result from various factors such as artificial intelligence, modular solutions, or organizational restructuring. Once established, progress is self-sustaining and can be replicated by others given similar contexts and resources, distinguishing it from mere coincidence or temporary gains.

Progress is not necessarily a linear movement towards an ultimate truth or goal, but it is an ongoing evolution of society and human knowledge, with open-ended objectives \cite{Kuhn1994structure}, \cite{Nisbet1994History}, \cite{Tam2024Progress}. Progress involves increasing complexity, specialization, and refinement of understanding in various fields of human endeavour toward some goal or ideal state \cite{Kuhn1994structure}, \cite{Tam2024Progress}, \cite{Bird2007What}, \cite{Niiniluoto1980Scientific}.

Additionally, the structure of human societies and institutions can foster or hinder progress. People are shaped by their environments and cultural paradigms to address specific challenges hence an organizations or communities can drive advancement in their specific areas, which also includes competition between ideas, methods, or systems that spur innovation and improvement, with new approaches or paradigms to solve pressing problems while building upon, rather than completely discarding, past achievements \cite{Kuhn1994structure}, \cite{Nisbet1994History}, \cite{Tam2024Progress}.

\section{How to Ensure Progress?}
We democratize how to make progress by:
\begin{enumerate}
    \item Embracing non-linear yet cumulative advancements, with emphasis on paradigm shifts \cite{Kuhn1994structure} and iterative processes mindset \cite{Brown2008Design}, \cite{Brown2011Change}, \cite{Kurek2023Sustainable}, \cite{Razzouk2012What}. Importantly, ensure that cumulative advancements are directed towards clear objectives to avoid aimless progression or Brownian motion.
    \item Understanding progress by its purpose and underlying mechanisms \cite{Kuhn1994structure} with a grasp of the historical context and philosophical underpinnings \cite{Nisbet1994History}. However, progress can only be ensured by working beyond the current understanding. Therefore, use this knowledge as a foundation to push boundaries and explore new frontiers.
    \item Collaborating with multi-stakeholders, as progress often occurs at disciplinary intersections \cite{Kuhn1994structure}. Engage stakeholders in decision-making to promote co-creation, ensure wide distribution of benefits, and consider marginalized perspectives \cite{Brown2008Design}, \cite{Kurek2023Sustainable}, \cite{Lu2023Progress}. Crucially, educate stakeholders on the current state of the art before soliciting their input, ensuring their perspectives build upon and extend existing knowledge.
    \item Experimenting and prototyping at the system level with proper methodologies for repeatability, considering interconnections between societal and environmental elements \cite{Kuhn1994structure}, \cite{Brown2008Design}, \cite{Brown2011Change}, \cite{Kurek2023Sustainable}, \cite{Razzouk2012What}. Ensure that these experiments are guided by clear end-to-end expectations of desired behaviors, with the system designed to meet these specific objectives.
    \item Recognizing that conflicting progress is not a problem, but rather a valuable tool for refinement. Embrace these conflicts as a competitive approach to confirm and strengthen the progress itself \cite{Kuhn1994structure}. This adversarial process can help identify weaknesses, validate assumptions, and ultimately lead to more robust and well-tested advancements.
\end{enumerate}

\section{Making Progress in the Post-AI Era}
As AI rapidly advances, we must recalibrate our approach to progress. The clock is ticking, and we cannot afford to be complacent. First, we need a more holistic, interdisciplinary perspective to navigate AI's complex impacts across with proper governance structures to keep pace with AI's evolution while providing stable foundations \cite{Bengio2024Managing}, \cite{Camilleri2024Artificial}, \cite{OzmenGaribay2023Six}, \cite{Singh2024Artificial}. Then, to leverage AI for progress, we consider multiple industries where AI can accelerate advances \cite{Bahoo2023Artificial}. Lastly, advancing AI itself requires a focus on safety, interpretability, energy efficiency, and more general intelligence with ethical considerations and societal impact assessment \cite{OzmenGaribay2023Six}, \cite{Boddington2023AI}, \cite{Bolte2024Sustainable}.

Yet there are several fundamental limitations of AI for making progress. AI tools may create illusions of understanding, where users believe they comprehend more than they do which could lead to overconfidence in AI-generated insights across various fields \cite{Messeri2024Artificial}, \cite{Rabb2019Individual}, \cite{Rozenblit2002misunderstood}. While several mitigations exist \cite{Buçinca2021To}, \cite{Jacovi2021Formalizing}, \cite{Marvin2024Prompt}, AI still risks fostering "monocultures of knowing," narrowing question diversity \cite{Messeri2024Artificial}. AI's big data approach prioritizes quantitative, reductive methods, potentially marginalizing nuanced, context-dependent knowledge, even reinforcing bias \cite{Messeri2024Artificial}, \cite{Barberá2015Birds}, \cite{Boyd2012CRITICAL}. Without proper role-playing and dataset awareness, AI might reduce problem-solving team diversity, paradoxically decreasing innovation \cite{Messeri2024Artificial}, \cite{AlShebli2018preeminence}, \cite{Hong2004Groups}, \cite{Nakadai2023AI}, \cite{Sulik2022Diversity}. Finally, AI-driven efficiency, coupled with publication pressure, may sacrifice deeper understanding for quantity, leading to "producing more but understanding less" \cite{Messeri2024Artificial}, \cite{Amutuhaire2022Reality}, \cite{Elbanna2023From}, \cite{Stokel-Walker2024Chatbot}, \cite{Wang2023Scientific}. Overall, mitigating those limitations is important to actively embrace AI's potential while vigilantly addressing its limitations, fostering diverse perspectives, and prioritizing deep understanding to ensure that our progress serves humanity.

\section{Conclusion}
Progress in the post-AI era demands an approach that balances technological advancement with human-centric values. It requires clear objectives, interdisciplinary collaboration, and a willingness to push beyond current understanding. AI offers unprecedented opportunities for advancement, yet its limitations confirm the continued importance of human creativity, consciousness, and abductive reasoning. True progress will only emerge from a symbiosis of AI capabilities and human ingenuity. The future is in our hands – let us create a world we can be proud to pass on to future generations.

\newpage
\bibliographystyle{unsrt}
\bibliography{references}

\end{document}